\documentclass{ws-procs975x65}
\usepackage{graphicx}
\def\beq{\begin{equation}}
\def\eeq{\end{equation}}
\newcommand{\be}{\begin{equation}}
\newcommand{\ee}{\end{equation}}
\newcommand{\ba}{\begin{eqnarray}}
\newcommand{\ea}{\end{eqnarray}}
\def\quarter{\textstyle{\frac{1}{4}}}
\def\half{\textstyle{\frac{1}{2}}}

\begin{document}

\title{Rotating black holes pierced by a cosmic string}

\author{David Kubiz{\v n}{\' a}k}

\address{Perimeter Institute, 31 Caroline Street North, Waterloo, ON, N2L 2Y5,
Canada\\
E-mail: dkubiznak@perimeterinstitute.ca}

\begin{abstract}
A rotating black hole threaded by an infinitely long cosmic string
is studied in the framework of the Abelian Higgs model.
We show that contrary to a common belief,
in the presence of rotation the backreaction of the string does not induce a simple conical deficit.
This leads to new distinct features of the Kerr--string system such as modified ISCO or shifted ergosphere, though
these effects are likely outside the range of observational precision.
For an extremal rotating black hole, the system exhibits a first-order
phase transition for the gravitational Meissner effect: whereas the horizon of large black holes is pierced by the vortex, small black holes exhibit a flux-expelled solution, with the gauge and scalar field remaining
identically in their false vacuum state on the event horizon.
A brief review prepared for the MG14 Proceedings.
%A generalization to the AdS case and its relevance for the AdS/CFT correspondence is also briefly mentioned.

\end{abstract}

%Sample file; \LaTeX;
%Prepared for MG14 Proceedings; World Scientific Publishing.}

\bodymatter

%%%%%%%%%%%%%%%%%%%%%%%%%%%%%%%
\section{Cosmic strings}
Cosmic strings are stable linear defects %with energy density equal to tension along its length
that potentially arise in a symmetry breaking phase transition in the early universe or in brane inflation models of string theory [\refcite{Vilenkin}].
The gravitational backreaction of an isolated string yields a conical deficit that generates a gravitational lens.
Perhaps surprisingly, cosmic strings can give rise to a long range hair that can be supported by the black hole event horizon: captured by a black hole, cosmic string hair provides a counter-example to the black hole {\em no hair theorem} [\refcite{AFV}, \refcite{AGK},
\refcite{BEG}, \refcite{ROG}, \refcite{GB}, \refcite{GKW}, \refcite{GMdS}, \refcite{VH}, \refcite{DGM1}, \refcite{Gregory:2014uca}].

The abelian Higgs model provides a simple and elegant model for a cosmic string that can capture all essential features: energy/tension balance, a finite width core of condensate, gauge flux through. The action reads\footnote{We use
units in which $\hbar=c=1$ and a mostly minus signature.}
\be \label{abhact}
S = \int d^4x \sqrt{-g} \left [-\frac{1}{16\pi G}R+ D_{\mu}\Phi ^{\dagger}D^{\mu}\Phi -
{\quarter} {\tilde F}_{\mu \nu}{\tilde F}^{\mu \nu} - {\quarter}\lambda
(\Phi ^{\dagger} \Phi - \eta ^2)^2 \right ]\,,
\ee
where $\Phi$ is the Higgs field, and $A_\mu$ the U(1) gauge
boson with field strength ${\tilde F}_{\mu\nu}$.
After a standard re-definition of field variables, see e.g. [\refcite{Gregory:2014uca}], the abelian Higgs equations of motion take the following form:
\ba \label{vorteqn}
\nabla _{\mu}\nabla ^{\mu} X - P_{\mu}P^{\mu}X + \frac{\lambda\eta^2}{2}
X(X^2 -1) &=& 0\,, \nonumber\\
\nabla _{\mu}F^{\mu \nu} + 2e^2\eta^2 X^2 P^{\nu}&=& 0\,.
\ea
Here, $X$ represents the residual massive Higgs field, $P_\mu$ is the
massive vector boson, and $F^{\mu\nu}$ its field strength.
One can set the units of energy so that $\sqrt{\lambda}\eta=1$, effectively stating that our Higgs field has order
unity mass. As per usual we have also introduced the Bogomol'nyi parameter
$\beta = \lambda/2e^2\,.$

\begin{figure}[h]
\begin{center}
\includegraphics[width=0.6\textwidth,height=0.25\textheight]{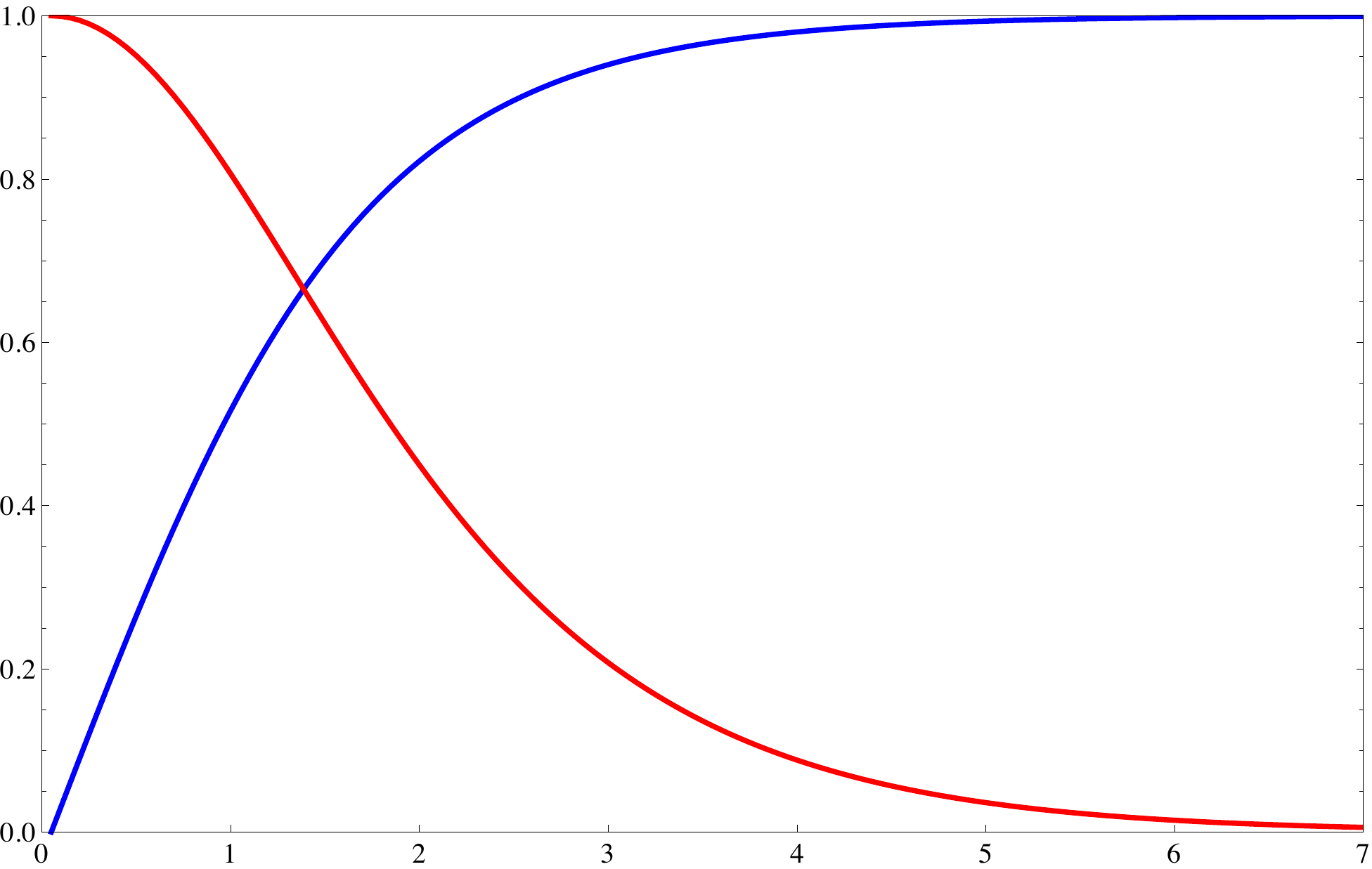}
\end{center}
\caption{{\bf Nielsen--Olesen vortex.} Numerical solution for $X_0$ (blue) and
$P_0$ (red) is displayed for $\beta=1$. %Notice $P_0$ falls off a little more slowly on this scale.
}
\label{fig:NOV}
\end{figure}
The `simplest solution' of the above equations is the
{\em Nielsen--Olesen vortex}, describing an infinitely long static cosmic string in a flat space [\refcite{NO}].
In cylindrical polar coordinates $(t, r,\phi,z)$, it reads
$X = X_0(R)\,, P= P_0(R)d\phi \,,$ where $R=r\sqrt{\lambda}\eta$, and
$X_0$ and $P_0$ satisfy
\be \label{basic}
X''_0 + \frac{X'_0}{R} - \frac{X_0P^2_0}{R^2}
- {\half} X_0(X^2_0-1) = 0\,, \quad
P''_0-\frac{P'_0}{R} - \frac{X^2_0P_0}{\beta} = 0\,.
\ee
The system can be easily solved numerically and the solution for $\beta=1$ is displayed in Fig.~\ref{fig:NOV}.
%It can be shown
%The backreaction of the vortex introduces a conical deficit \cite{AFV,}

To find the backreaction of the vortex, we can proceed perturbatively [\refcite{AGK}], performing an expansion
in a dimensionless ratio  $\epsilon\equiv 8\pi G \eta^2,$
%The gravitational effect of the vortex is determined by $\epsilon\equiv 8\pi G_N \eta^2\,,$ a dimensionless ratio
which is typically of order $10^{-7}-10^{-12}$ for cosmic strings of cosmological relevance. That is,
the first-order perturbation of the metric is determined from
\be\label{Rmunu}
R_{\mu\nu}=\epsilon \Bigl(T_{\mu\nu}-\frac{1}{2}Tg_{\mu\nu}\Bigr)\,,
\ee
where $T_{\mu\nu}$ stands for the asymptotic form of the stress-energy tensor of the Nielsen--Olesen vortex.
By performing this calculation, one finds  that the effect of the vortex is indeed to introduce a conical deficit, with the deficit angle
given by
\be
\delta \phi=2\pi\epsilon \mu\,,\quad  \mbox{where} \quad \mu=\int_0^\infty R{\cal E}dR
\ee
is the energy per unit length of the string, and we have denoted ${\cal E}=T^t_t$.

%%%%%%%%%%%%%%%%%%%%%%%%%%%%%%%%%%%%%%%%%%%%%%%%%
\begin{figure}[h]
\begin{center}
\includegraphics[width=0.32\textwidth,height=0.26\textheight]{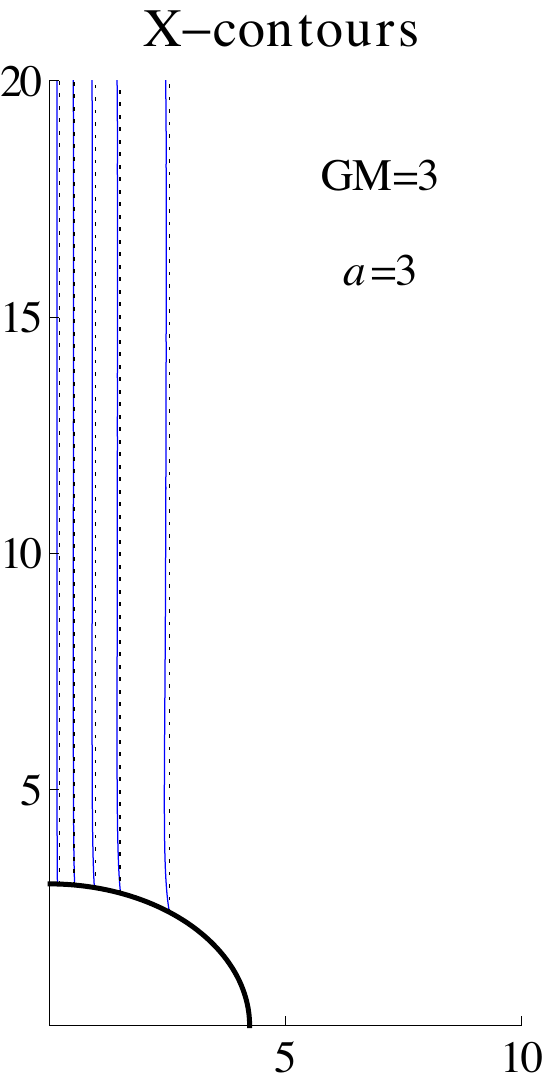}
\includegraphics[width=0.32\textwidth,height=0.26\textheight]{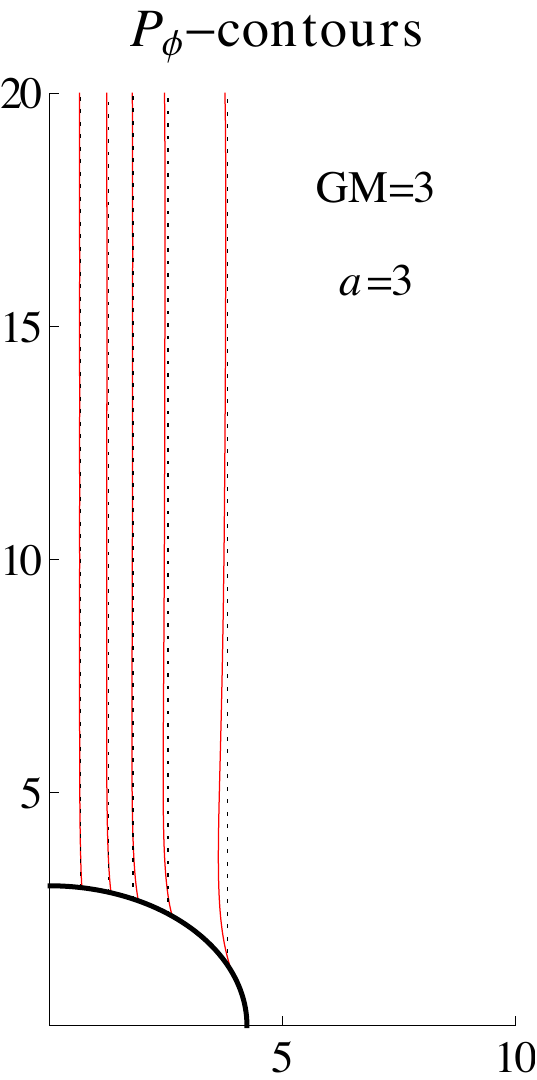}
\includegraphics[width=0.32\textwidth,height=0.26\textheight]{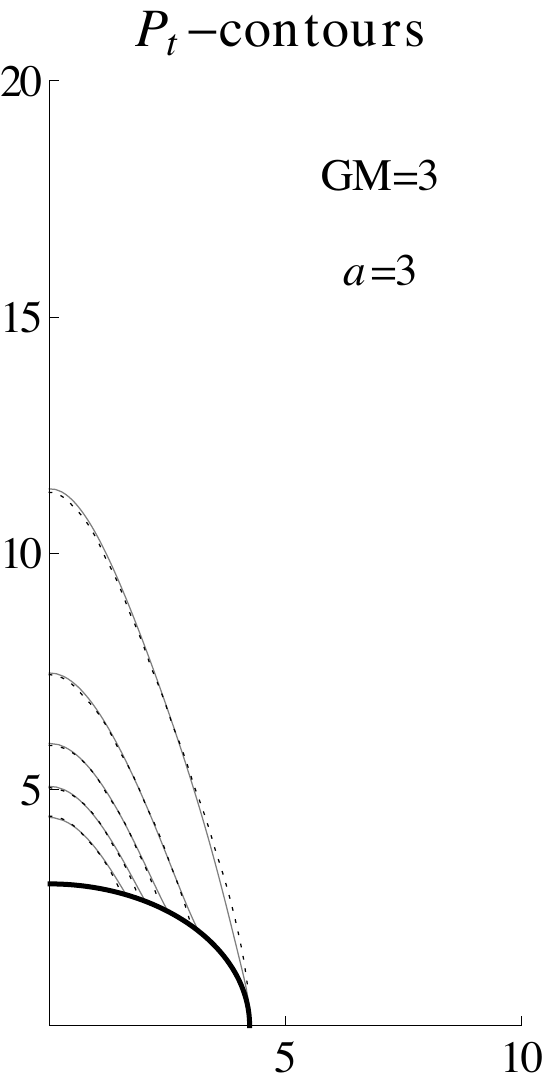}
\end{center}
\caption{{\bf Vortex fields.}
A comparison of the approximate and exact numerical solutions
for an extremal $GM=a=3$ Kerr black hole.
%In spite of the low
%value of black hole mass, \eqref{ApproxSol} is still an
%extremely good approximation to the actual result. Here,
The Higgs contours are in blue, the $P_\phi$ contours in red,
the $P_t$ contours in grey, and all the corresponding approximate
solution contours in dashed black. Contours are shown for $X,P_\phi
=0.1, 0.3, 0.5, 0.7, 0.9$, and for $P_t = -0.099, -0.077,
-0.055, -0.033, -0.011$.}
\label{aba:fig1}
\end{figure}

%%%%%%%%%%%%%%%%%%%%%%%%%%%%%%%%%%%%%%%%%
\section{The Kerr--string system}
To find an approximate solution for a cosmic string captured by a rotating black hole we proceed as follows.\footnote{We consider a `final equilibrium state' of the black hole--cosmic string  interacting system where an infinite cosmic string is aligned along the axis of rotation of the black hole.}
One first obtains an analytic approximation for the vortex in a fixed Kerr background and then backreacts
its energy-momentum to obtain the metric valid to the leading order in $\epsilon$, using \eqref{Rmunu}.
The asymptotic form of the resulting approximate solution then reads
%\footnote{In a given approximation, one can also write the
%following formula for the vector potential: $P=P_0(R)\bigl[d\phi-\frac{2aGMr}{\Gamma}dt\bigr]\,.$}  
[\refcite{GKW}]
\ba\label{sol}
ds^2&\simeq&-\frac{\Delta\Sigma}{\Gamma}dt^2+(1-2\epsilon \mu)\frac{\sin^2\!\theta \Gamma}{\Sigma}\Bigl[d\phi+\frac{2aGMr}{\Gamma}dt\Bigr]^2
+\Sigma\Bigl(\frac{dr^2}{\Delta}+d\theta^2\Bigr)\,,\nonumber\\
X &\simeq& X_0(R)\,,\quad
P\simeq P_0(R)\Bigl[d\phi-\frac{2aGMr}{\rho^4} dt\Bigr]\simeq P_0(R)\Bigl[d\phi-\frac{2aGMr}{\Gamma} dt\Bigr]\,,
\ea
%\ba\label{sol}
%ds^2&\simeq&-\frac{\Delta\Sigma}{\Gamma}dt^2+(1-2\epsilon \mu)\frac{\sin^2\!\theta \Gamma}{\Sigma}\Bigl[d\phi+\frac{2aGMr}{\Gamma}dt\Bigr]^2
%+\Sigma\Bigl(\frac{dr^2}{\Delta}+d\theta^2\Bigr)\,,\nonumber\\
%X &\simeq& X_0(R)\,,\quad
%P_\phi \simeq P_0(R)\,,\quad
%P_t \simeq -\frac{2aGMr}{\rho^4} P_0(R)\,,
%\ea
where $X_0$ and $P_0$ stand for the Nielsen--Olesen functions obeying \eqref{basic}, and
\ba
\Sigma&=&r^2+a^2\cos^2\!\theta\,,\quad \Delta=\rho^2-2GMr\,,\quad
\Gamma = \rho^4 - \Delta a^2 \sin^2\theta\,,\nonumber\\
\rho^2&=&r^2+a^2\,,\quad R=\rho\sin\theta\,.
\ea
A generalization to the charged and rotating AdS case is rather
straightforward and can be found in [\refcite{Gregory:2014uca}].

Concentrating first on the vortex fields, we note the presence of the time-component of the gauge potential, reflecting the fact that
the rotating geometry of Kerr induces a near horizon electric field. Fig.~\ref{aba:fig1} illustrates how good an approximation to the full numerical solution the expression for the vortex fields is in the case of
extremal black hole. In general, the approximation is valid for thin (relative to black hole) strings.

Let us next look at the metric. Obviously, in the absence of string, $\mu=0$,  we recover the standard Kerr geometry written in Boyer--Lindquist coordinates. For $\mu\neq 0$, the metric describes a spacetime `outside of the string' and solves the vacuum Einstein equations to  a given order of approximation. % (it is not exactly a vacuum spacetime with $R_{\mu\nu}=0$).
Clearly, in the presence of rotation the backreaction of the string {\em does not induce a simple conical deficit}.
In fact, the angular deficit approaches the standard conical deficit only at large distances, while
in the vicinity of the black hole, as far as the Boyer--Lindquist
coordinates are concerned, the deficit is felt not only by the angular
coordinate $\phi$, but also by the time component of the metric.
In particular, on the horizon and upon a transformation to
a frame co-rotating with the black hole, $\tilde \phi = \phi -
\Omega_+ t$, we realize that the effect of the cosmic string
is to again remove a deficit angle, however now with respect to
a local co-rotating frame. In the intermediate region the backreaction of the string seems (in a given approximation) rather `complicated' and depends on both $\theta$ and $\phi$: it is the 1-form $d\phi+(2aGMr/\Gamma) dt$ that `feels' the conical deficit.

\begin{figure}[h]
\begin{center}
\includegraphics[width=0.6\textwidth,height=0.26\textheight]{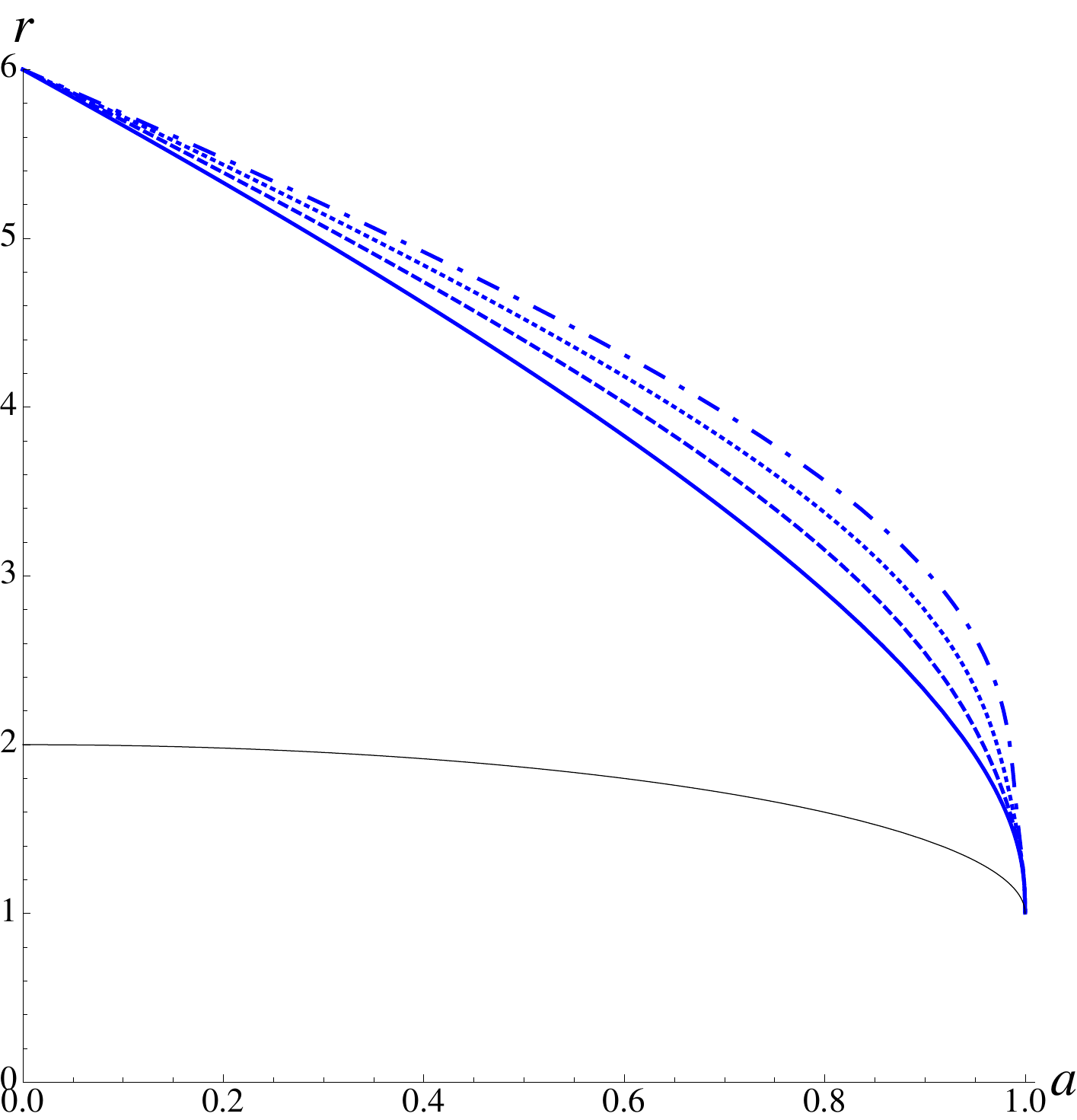}
\end{center}
\caption{%{\bf New physical features.} {\em Left.}
{\bf Shifting of ICSO's} as a function of $a$ (measured in
units of $GM$) for $\epsilon=0, 0.1, 0.2, 0.3$, represented by
solid, dashed, dotted, and dot-dashed blue lines respectively.
The black line indicates the location of the event horizon.
%{\em Right.} Gravitational lensing in equatorial plane for impact parameters $b=5, 10$.
}
\label{fig:isco}
\end{figure}

In other words, cosmic string seems to induce a conical deficit  ``from a perspective of the black hole'' rather than with respect to the static frame at infinity as assumed in all previous works. This has several consequences. For example, the original Kerr geometry admits hidden symmetries of Killing and Killing--Yano tensors that render the geodesic motion completely integrable, e.g. [\refcite{Frolov:2008jr}]. When the string is present, these symmetries no longer exist and the geodesic motion will be generically chaotic, we no longer have Carter's like constant of motion.
A variety of new physical phenomena emerges, such as: loss of integrability of geodesic motion, altered gravitational lensing, modified  ergosphere, and shifted ISCO's, see e.g. Fig.~\ref{fig:isco}.
%It would be really interesting to find numerically a full backreacted solution and study these effects in greater detail.

%%%%%%%%%%%%%%%%%%%%%%%%%%%%%%%%%%%%%%%%%%%%%%
\section{Gravitational Meissner effect}
It has been observed by Bi{\v c}{\'a}k and Dvo{\v r}{\'a}k long time ago [\refcite{Expulsion2}, \refcite{Bicak2}]
that {\em extremal black holes} tend to expel stationary (axisymmetric) electromagnetic fields, see also [\refcite{Expulsion1}, \refcite{Expulsion3}, \refcite{Penna:2014aza}, \refcite{Bicak:2015lxa}]. This so called gravitational Meissner effect has been speculated
to be able to quench the power of astrophysical jets as the black hole is `spun up' towards extremality.  [Expelled magnetic fields no longer pierce the horizon and the mechanism for extraction of black hole rotational energy terminates.]

\begin{figure}[h]
\begin{center}
\includegraphics[width=0.47\textwidth,height=0.26\textheight]{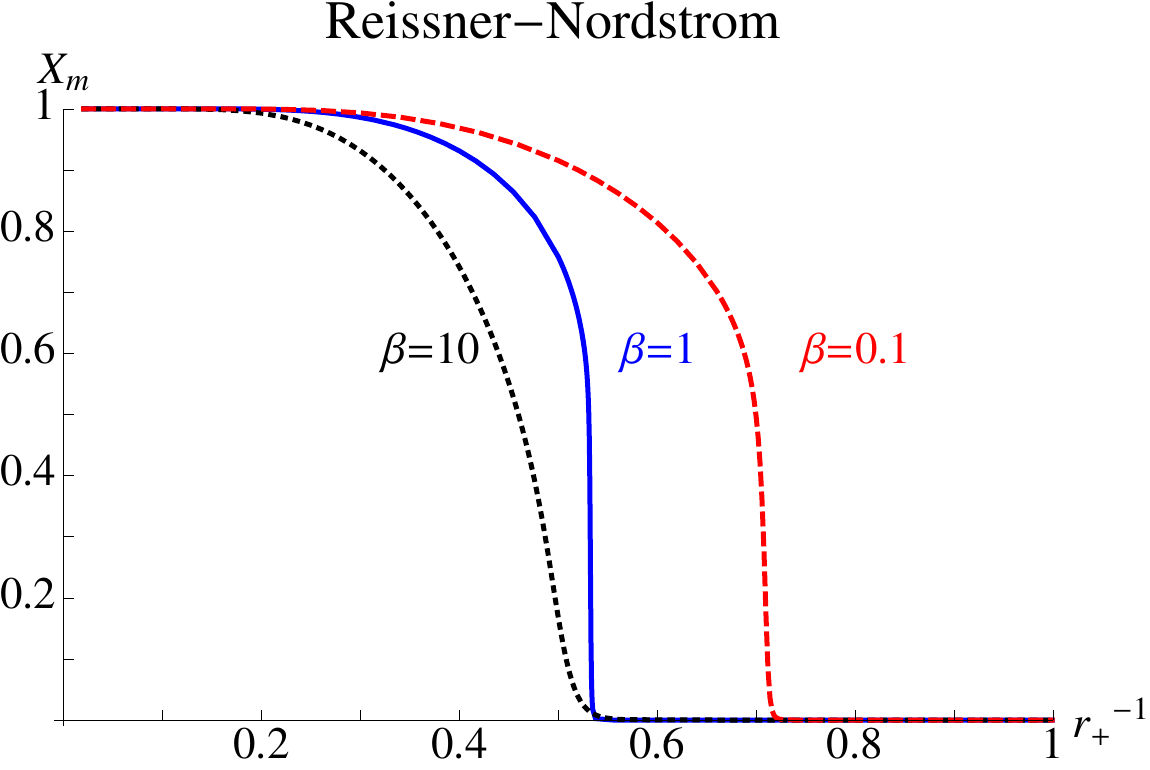}\nobreak
\includegraphics[width=0.47\textwidth,height=0.26\textheight]{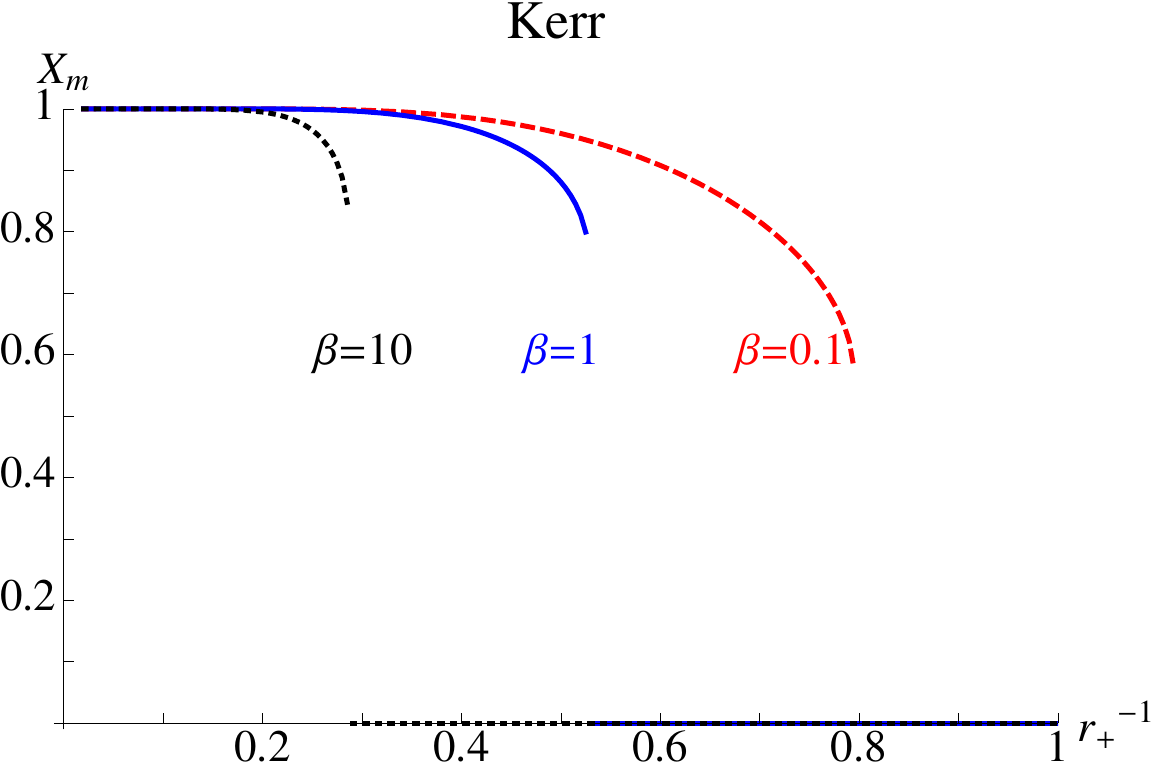}
\end{center}
\caption{{\bf Gravitational Meissner effect: black hole--vortex system.} The maximum value of the Higgs field on the horizon, $X_m$, is
plotted against the horizon radius $r_+$ for the extremal charged (left) and the extremal rotating (right) black holes.
In the first case we observe a second-order phase transition between penetration and expulsion, whereas the analogous phase transition for the rotating black hole seems of the second-order. This phenomenon is shown for different values of the Bogomolnyi parameter $\beta$.
%: $\beta=10$ in
%dotted black, $\beta=1$ in solid blue, and $\beta=0.1$ in
%dashed red.
}
\label{fig:RNK}
\end{figure}

For the cosmic string vortex the gravitational Meissner effect is even more interesting (at least in the  test field approximation in which it has been studied)
[\refcite{BEG}, \refcite{GKW}, \refcite{Gregory:2014uca}]. Namely, for a given extremal black hole there exists a {\em critical radius} of the string: a thin string penetrates the horizon
whereas a large one will be expelled. Surprisingly, the phenomenon seems slightly different for extremal charged and extremal rotating black holes:
in the charged case we observe a second-order phase transition between expulsion and penatration, whereas the analogous phase transition in the rotating case seems of the first-order, see Fig.~\ref{fig:RNK}. We remark that although the precise value of the critical radius has to be determined numerically, its existence can be predicted analytically, see  [\refcite{BEG}, \refcite{GKW}, \refcite{Gregory:2014uca}] for more details.

\section{Conclusions}
Let us summarize the main results of this brief review. We have shown that a generic (rotating) black hole can sport long hair of an abelian Higgs vortex. This leads to an electric field in the polar regions of the black hole. For extremal black holes, dependent on whether the black hole is charged or rotating, we observed a second-order or first-order phase transition for the gravitational Meissner effect. %It would be interesting
Especially interesting is the backreaction of the vortex. Up to now all previous works assumed a simple conical deficit with respect to the observer at infinity. We solved for a first order correction to the Kerr geometry using the stress-energy tensor for an approximate solution
and shown that the effect of the string is to remove a deficit angle from a ``perspective of the black hole''.
This leads to new (perhaps potentially observable) features of the black hole--cosmic string system.
It would be interesting to fully numerically integrate the Einstein--abelian Higgs equations. This would allow one to find the full backreacted metric and study potential astrophysical applications.

\section*{Acknowledgments}
This research was supported in part by Perimeter Institute for Theoretical Physics and by the Natural Sciences and Engineering Research Council of Canada. Research at Perimeter Institute is supported by the Government of Canada through Industry Canada and by the Province of Ontario through the Ministry of Research and Innovation.


\begin{thebibliography}{0}

\bibitem{Vilenkin}
%\cite{Vilenkin:1984ib}
A.~Vilenkin and E.~P.~S. Shellard,
{\em Cosmic Strings and other Topological Defects}.
\newblock Cambridge University Press, Cambridge, England, 1994.


\bibitem{AFV}
%\cite{Aryal:1986sz}
M.~Aryal, L.~Ford, and A.~Vilenkin,
{\it {Cosmic strings and black holes}},
Phys.\ Rev.\  {\bf D34} (1986) 2263.

\bibitem{AGK}
%\cite{Achucarro:1995nu}
A.~Achucarro, R.~Gregory, and K.~Kuijken,
{\it {Abelian Higgs hair for black holes}},
Phys.\ Rev.\  {\bf D52} (1995) 5729--5742,
%[\href{http://xxx.lanl.gov/abs/gr-qc/9505039}
[{{\tt gr-qc/9505039}}].

\bibitem{BEG}
%\cite{Bonjour:1998rf}
F.~Bonjour, R.~Emparan, and R.~Gregory,
{\it {Vortices and extreme black holes: The Question of flux expulsion}},
Phys.\ Rev.\  {\bf D59} (1999) 084022,
%[\href{http://xxx.lanl.gov/abs/gr-qc/9810061}
[{{\tt gr-qc/9810061}}].

\bibitem{ROG}
L.~Nakonieczny and M.~Rogatko,
{\it Abelian-Higgs hair on stationary axisymmetric black hole
in Einstein-Maxwell-axion-dilaton gravity},
Phys.~Rev.~ {\bf D88}, 084039 (2013),
%[\href{http://arXiv.org/abs/arXiv:1310.5929}
[{{\tt arXiv:1310.5929 [hep-th]}}].

\bibitem{GB}
%\cite{Ghezelbash:2001pq}
A.~Ghezelbash and R.~B. Mann,
{\it {Abelian Higgs hair for rotating and charged black holes}},
Phys.\ Rev.\  {\bf D65} (2002) 124022,
%[\href{http://xxx.lanl.gov/abs/hep-th/0110001}
[{{\tt hep-th/0110001}}].

\bibitem{GKW}
%\cite{Gregory:2013xca}
R.~Gregory, D.~Kubiznak and D.~Wills,
{\it Rotating black hole hair,}
JHEP {\bf 1306}, 023 (2013),
%\href{http://xxx.lanl.gov/abs/1303.0519}
[{{\tt arXiv:1303.0519 [gr-qc]}}].

\bibitem{GMdS}
%\cite{Ghezelbash:2002cc}
A.~Ghezelbash and R.~Mann,
{\it {Vortices in de Sitter space-times}},
{\em Phys.Lett.} {\bf B537} (2002) 329--339,
%[\href{http://xxx.lanl.gov/abs/hep-th/0203003}
[{{\tt hep-th/0203003}}].

\bibitem{VH}
%\cite{oai:arXiv.org:hep-th/0105134}
M.~H.~Dehghani, A.~M.~Ghezelbash and R.~B.~Mann,
{\it Vortex holography,}
Nucl.\ Phys.\ {\bf B625}, 389 (2002),
%[\href{http://xxx.lanl.gov/abs/hep-th/0105134}
[{{\tt hep-th/0105134}}].

\bibitem{DGM1}
%\bibitem{DGM2}
%\cite{Dehghani:2001nz}
M.~Dehghani, A.~Ghezelbash, and R.~B. Mann,
{\it {Abelian Higgs hair for AdS-Schwarzschild black hole}},
Phys.\ Rev.\  {\bf D65} (2002) 044010,
%[\href{http://xxx.lanl.gov/abs/hep-th/0107224}
[{{\tt hep-th/0107224}}].


%\cite{Gregory:2014uca}
\bibitem{Gregory:2014uca}
  R.~Gregory, P.~C.~Gustainis, D.~Kubiznak, R.~B.~Mann and D.~Wills,
{\it Vortex hair on AdS black holes},
  JHEP {\bf 1411}, 010 (2014),
 % doi:10.1007/JHEP11(2014)010
  [{\tt arXiv:1405.6507 [hep-th]}].
  %%CITATION = doi:10.1007/JHEP11(2014)010;%%
  %1 citations counted in INSPIRE as of 25 Dec 2015

\bibitem{NO}
%\cite{Nielsen:1973cs}
H.~B. Nielsen and P.~Olesen,
{\it {Vortex Line Models for Dual Strings}},
Nucl.\ Phys.\  {\bf B61} (1973) 45--61.


%\cite{Frolov:2008jr}
\bibitem{Frolov:2008jr}
  V.~P.~Frolov and D.~Kubiznak,
  {\it Higher-Dimensional Black Holes: Hidden Symmetries and Separation of Variables},
  Class.\ Quant.\ Grav.\  {\bf 25}, 154005 (2008),
  %doi:10.1088/0264-9381/25/15/154005
  [{\tt arXiv:0802.0322 [hep-th]}].
  %%CITATION = doi:10.1088/0264-9381/25/15/154005;%%
  %80 citations counted in INSPIRE as of 25 Dec 2015


\bibitem{Expulsion2} J.~Bi\v{c}\'ak and L. Dvo\v{r}\'ak,
{\it Stationary electromagnetic fields around black holes.
II. General solutions and the fields of some special sources
near a Kerr black hole}.
General Relativity and Gravitation, {\bf 7} 959 (1976).

\bibitem{Bicak2}
J.~Bi\v{c}\'ak and L. Dvo\v{r}\'ak,
{\it Stationary electromagnetic fields around black holes. III.
General solutions and the fields of current loops near the
Reissner-Nordstr\"om black hole},
Phys. Rev. {\bf D22} 2933 (1980).

\bibitem{Expulsion1}
A.~R.~King, J.~P.~Lasota, and W.~Kundt,
{\it Black holes and magnetic fields},
Phys. Rev. {\bf D12} 3037 (1975).


\bibitem{Expulsion3}
A.~Chamblin, R.~Emparan and G.~W.~Gibbons,
{\it Superconducting p-branes and extremal black holes},
Phys.~Rev.~{\bf D58}, 084009 (1998),
%[\href{http://arXiv.org/abs/arXiv:9806017}
[{{\tt arXiv:9806017 [hep-th]}}].


%\cite{Penna:2014aza}
\bibitem{Penna:2014aza}
  R.~F.~Penna,
{\it Black hole Meissner effect and Blandford-Znajek jets},
  Phys.\ Rev.\ D {\bf 89}, no. 10, 104057 (2014),
  %doi:10.1103/PhysRevD.89.104057
  [{\tt arXiv:1403.0938 [astro-ph.HE]}].
  %%CITATION = doi:10.1103/PhysRevD.89.104057;%%
  %10 citations counted in INSPIRE as of 25 Dec 2015


%\cite{Bicak:2015lxa}
\bibitem{Bicak:2015lxa}
  J.~Bi{\v c}{\'a}k and F.~Hejda,
{\it Near-horizon description of extremal magnetized stationary black holes and Meissner effect},
  Phys.\ Rev.\ D {\bf 92}, %no. 10,
  104006 (2015),
  %doi:10.1103/PhysRevD.92.104006
  [{\tt arXiv:1510.01911 [gr-qc]}].
  %%CITATION = doi:10.1103/PhysRevD.92.104006;%%



\end{thebibliography}
\end{document}